\documentclass[aps,showpacs,pre,superscriptaddress,nofootinbib,notitlepage,a4paper]{revtex4}
\usepackage[english]{babel}
\usepackage{dsfont}
\usepackage{mathtools,nccmath}
\usepackage{amsmath}
\usepackage{amsfonts}
\usepackage{amssymb}
\usepackage{xcolor}
\usepackage{graphicx}
\usepackage{psfrag}
\usepackage{float}
\usepackage{placeins}
\usepackage{scalerel}

\usepackage{tikz}
\usetikzlibrary{patterns,decorations,arrows,shapes.geometric,arrows,mindmap,backgrounds,positioning,positioning,fit,backgrounds,positioning,arrows,matrix,calc}
\usetikzlibrary{shapes,backgrounds,mindmap,shadows,shapes.geometric,topaths,snakes,arrows,pgfplots.groupplots,fadings,calc,decorations.markings,positioning,chains,fit}
\usetikzlibrary{arrows,shapes,positioning,quotes}
\usetikzlibrary{decorations.markings}
\tikzstyle arrowstyle=[scale=1]
\tikzstyle directed=[postaction={decorate,decoration={markings,
    mark=at position .65 with {\arrow[arrowstyle]{stealth}}}}]
\usetikzlibrary{shapes.geometric}
\tikzset{mynode/.style={draw, shape=circle, fill=black, color=black, line width=0pt, inner sep=2.5pt, label position=0, text=black}}
\tikzset{myedge/.style={draw,thick,solid}}
\tikzset{bicoloredge/.style={dashed, dash pattern=on 4pt off 4pt, blue, postaction={draw, dashed, dash pattern=on 4pt off 4pt, red, dash phase=4pt}, ultra thick}}
\usepackage{circuitikz}
\usepackage{ifthen}
\usepackage[hidelinks]{hyperref}
\usepackage{adjustbox}

\newcommand{\beq}{\begin{equation}} 
\newcommand{\eeq}{\end{equation}}
\newcommand{\bea}{\begin{align}}
\newcommand{\eea}{\end{align}}

\newcommand{\E}{\mathbb{E}}

\newcommand{\one}{\mathds{1}}
\newcommand{\hS}{\widehat{S}}
\newcommand{\R}{\mathbb{R}}
\newcommand{\eqd}{\overset{\rm d}{=}}
\newcommand{\oeta}{\overline{\eta}}
\newcommand{\s}{\sigma}

\newcommand{\la}{\langle}
\newcommand{\ra}{\rangle}
\newcommand{\A}{\mathcal{A}}
\newcommand{\M}{\mathcal{M}}
\newcommand{\cR}{\mathcal{R}}
\newcommand{\Tr}{{\rm Tr}}
\newcommand{\hr}{\widehat{r}}
\newcommand{\hX}{\widehat{X}}
\newcommand{\htheta}{\widehat{\theta}}

\newcommand{\pzero}[2]{\ifthenelse{\equal{#2}{}}{\mathbb{P}_0^{(#1)}}{\mathbb{P}_0^{(#1)}[#2]}}
\newcommand{\pone}[2]{\ifthenelse{\equal{#2}{}}{\mathbb{P}_1^{(#1)}}{\mathbb{P}_1^{(#1)}[#2]}}

\begin{document}

\bibliographystyle{myunsrt}

\title{Some observations on the ambivalent role of symmetries \\ in Bayesian inference problems}

\author{Guilhem Semerjian}
\affiliation{Laboratoire de physique de l'\'Ecole Normale Sup\'erieure, ENS, Universit\'e PSL, CNRS, Sorbonne Universit\'e, Universit\'e Paris Cit\'e, F-75005 Paris, France} 

\begin{abstract}
We collect in this note some observations on the role of symmetries in Bayesian inference problems, that can be useful or detrimental depending on the way they act on the signal and on the observations. We emphasize in particular the need to gauge away unobservable invariances in the definition of a distance between a signal and its estimator, and the consequences this implies for the statistical mechanics treatment of such models, taking as a motivating example the extensive rank matrix factorization problem.

\end{abstract}

\maketitle

\section{Introduction}

The notion of symmetry is a fascinating concept with a strong aesthetical appeal that influenced several branches of science; for instance in mathematics it led to the development of representation and invariant theories~\cite{FultonHarris_book,GoWa_book}, while modern physics, and in particular high-energy physics, used it as a guiding principle to build the standard model of particle physics~\cite{Zee_book}. It also played a role in statistical mechanics, through notably the classification of phase transitions according to the symmetry group of the order parameters, and its spontaneous breaking. In the context of disordered systems G\'erard Toulouse, to whom this article is dedicated, formulated his views on glasses and spin-glasses in this group-theoretic language~\cite{Toulouse77,To79}. The focus of this note is on the effects of symmetries in inference and machine learning problems, and in particular on their statistical mechanics treatment. This topic has a long history rooted in classical statistics~\cite{Ea89,Wi90,LehmannCasella_book} and in the early days of the development of neural networks~\cite{ShTa89}, and has seen a more recent revival triggering a burst of activity on the role of invariances in modern architectures of machine learning~\cite{BrBrCoVe21,ViHoYaKeSc24}. These invariances can be hard-wired in the form of equivariant networks~\cite{CoWe16} (generalizing the convolutional networks that are translationally invariant to arbitrary symmetries), or exploited to perform ``data augmentation''~\cite{ChDoLe20}, namely creating artificial new samples by applying random transformations respecting the assumed invariance to observed data points. The consequences of symmetries on the loss landscape of artificial neural networks have also been studied, see for instance~\cite{SiGeJaSpHoGeBr21}.

We shall not attempt in this note an exhaustive review of this rich and growing field of research, but instead concentrate on a specific setting, namely the Bayesian inference problems and their statistical mechanics treatment, and collect a series of simple observations on the role symmetries plays in this context. We will argue that this role can be either beneficial or detrimental depending, roughly speaking, on the relative strength of the action of the symmetry on the quantity to be inferred and on the one which is observed. A particular emphasis shall be put on the negative side of this dichotomy, where the non-trivial unobservable invariance of the signal imposes to reconsider the definition of the distance between the signal and its estimation. This point was of course noticed in previous works, but we believe its consequences may have been underappreciated, in particular for the extensive-rank matrix factorization problem whose invariance group is growing with the system size.

The rest of the manuscript is organized as follows. In Sec.~\ref{sec_def} we define some notions related to Bayesian inference, to symmetries, and to the interplay between the two. Sec.~\ref{sec_asset} is then devoted to situations where symmetries are useful in the inference process, while Sec.~\ref{sec_nuisance} discuss the opposite case where they are detrimental. Further perspectives on possible ways to handle these difficulties are given in the conclusions of Sec.~\ref{sec_conclu}.

\section{Definitions}
\label{sec_def}

\subsection{Bayesian inference}
\label{sec_def_inference}

In an inference problem a signal of interest, denoted $S$, cannot be measured directly but has to be determined from indirect observations, called $Y$ in the following. In general $S$ and $Y$ can have a rich structure, and should be thought of as vectors of arbitrary dimensions. The Bayesian perspective on inference treats $S$ and $Y$ as random variables; one way to describe their distributions is as follows. The signal $S$ is drawn from a prior law $P_S(s) = \mathbb{P}(S=s)$ (we follow here the convention of denoting random variables with uppercase symbols, the values of their realizations with the corresponding lowercase letters, and assume for notational simplicity that the random variables take discrete values, we will later generalize to continuous ones). The observation $Y$ is then drawn conditionally on the signal, with the probability $P_{Y|S}(y|s)$, which can model the addition of noise to the true signal, or the flow of information contained in $S$ through an imperfect communication channel. From a probabilistic point of view it is however equivalent, and sometimes more convenient, to think of the pair of variables $(S,Y)$ as drawn with the joint probability law $P_{S,Y}(s,y)=P_S(s) P_{Y|S}(y|s)$. 

The goal of the observer is to propose an estimate $\hS(y)$ of the true value of the signal, based on the observation of the event $Y=y$, and assuming the knowledge of the probability laws underlying this generation process, namely the prior distribution $P_S$ and the noisy channel $P_{Y|S}$. In general the observer cannot be sure of the hidden value of the signal, and can only assign an a posteriori probability to its possible values; according to the Bayes theorem this posterior distribution $P_{S|Y}$ reads
\beq
P_{S|Y}(s|y) = \frac{P_S(s) P_{Y|S}(y|s)}{P_{Y}(y)} \ ,
\label{eq_posterior}
\eeq
on which the computation of $\hS(y)$ has to be based.

The tools of statistical mechanics of disordered systems have been applied to a wide variety of such problems~\cite{NishimoriBook01,Engel_book01,MezardMontanari07,ZdKr16_review,farbeyond}. As a matter of fact the probability law (\ref{eq_posterior}) can be viewed as the Gibbs-Boltzmann distribution for thermal variables $s$, with an Hamiltonian $H_y(s) = - \ln P_S(s) - \ln P_{Y|S}(y|s)$, where $y$ plays the role of a quenched disorder (with a so-called ``planted'' distribution), at inverse temperature $\beta=1$. In this perspective the denominator in (\ref{eq_posterior}) is viewed as a partition function ensuring the normalization of the law, $P_Y(y)=Z(y) = \sum_s P_S(s) P_{Y|S}(y|s)$, and one defines the quenched free-energy as $\E[\ln Z(Y)]$ (which is the opposite of the Shannon entropy of $Y$ and can be related, up to a simple additive constant, to the mutual information between $S$ and $Y$).

Let us present a few examples that will be used for illustration in the forthcoming sections:
\begin{itemize}

\item the stochastic block model (SBM) is an ensemble of random graphs that was introduced to study complex networks with a community structure~\cite{fienberg1981categorical,holland1983stochastic,BoJaRi07}: its nodes (or vertices) belong to ``communities'' and the probability of existence of an edge between a pair of nodes (modelling for instance a social interaction between two individuals) depends on the communities of the two nodes. In this context the signal $S$ corresponds to the communities of each node, and the observation $Y$ to the graph structure (i.e. the edges connecting some nodes), and the inference goal is to determine the hidden community structure from the knowledge of the graph. More precisely, denoting $N$ the number of vertices and $q$ the number of communities, one has $S=(\s_1,\dots,\s_N)$ where $\s_i \in \{1,\dots,q\}$ is a label specifying the community of the $i$-th node. The prior law $P_S$ is a product measure, each $\s_i$ is drawn independently at random from the probability law $\oeta=(\oeta_1,\dots,\oeta_q)$. The observed graph $Y$ is then generated by including each of the $N(N-1)/2$ possible edges $\la i,j \ra$ between pairs of distinct vertices with probability $c_{\s_i,\s_j}/N$, where $c$ is a $q \times q$ matrix with non-negative real elements $c_{\s,\tau}$, the latter measuring the ``affinity'' between individuals of communities $\s$ and $\tau$, and hence their tendency to create bonds between them. This problem was studied with statistical mechanics techniques in~\cite{Decelle_SBM_long}, unveiling phase transitions in the thermodynamic (large size) limit $N \to \infty$: depending on the parameters $\oeta$ and $c$ it might be possible or not to infer $S$ from $Y$ with an accuracy higher than a guess from the prior probability. Moreover the computational cost of this task can go from polynomial to exponential in $N$, opening the possibility of a phenomenon known as a statistical-computational gap when the algorithmic and information-theoretical phase transitions do not coincide. Parts of this scenario were proven rigorously later on, see~\cite{Cris_review_SBM,Abbe_review_SBM} for reviews of this research direction.

\item in matrix denoising problems the signal and observation $S$ and $Y$ are both matrices; in the simplest version they are real and symmetric, of size $n \times n$, and the observation channel corresponds to the addition of an independent noise matrix $Z$, also symmetric, namely $Y = S+Z$. The matrix character of the problem shows up in the structure of the laws of $S$ and $Z$, which for many natural ensembles of Random Matrix Theory (RMT), and most famously the Gaussian Orthogonal Ensemble (GOE), enjoy some invariance properties, for instance under orthogonal transformations. This inference problem was in particular studied in the seminal paper~\cite{BuAlBoPo16}, and revisited more recently in~\cite{MaKrMeZd22,BaMa22,TrErKrMaZd22,PoBaMa23,Se24,BaCaKoOk24}. 

\item a related problem is the matrix factorization one, where one observes a matrix $Y = XX^T +Z$, with $X$ an $n \times r$ matrix and $Z$ a noise, and the goal is to infer $X$ from $Y$ (to avoid confusion with the matrix denoising problem the signal is denoted $X$ in this case). The matrix denoising problem can thus be seen as a simplified version of the matrix factorization one, in which one only aims at reconstructing the product of the two unknown factors (i.e. the Wishart matrix $XX^T$) instead of the factor $X$ individually. Matrix factorization appears in various problems of machine learning~\cite{OlFi96,ZiPe01,KrMuRaEnLeSe03,SpWaWr12}, and has been the subject of an intense research activity. The low-rank regime of the matrix factorization problem, when $r$ remains finite in the large $n$ limit, is now rather well understood~\cite{RaFl12,BaDiMaKrLeZd16,LeKrZd17,LeMi19,BaMa19,MoWe22}, with a precise characterization of the information-theoretical optimal accuracy of estimation, as well as the performances of polynomial time algorithms for this task. By contrast the extensive rank regime, with $r$ proportional to $n$ in the large size limit, presents additional challenges that are not yet all resolved. The initial proposal of~\cite{SaKa13,KaKrMeSaZd16} was later unveiled to be incorrect~\cite{Sc18}, and despite a series of recent efforts~\cite{MaKrMeZd22,BaMa22,TrErKrMaZd22,PoBaMa23,BaKoRa24,PoMa23,PoMa23b,LaMeGa23,CaMe23a,CaMe23b} the current understanding of this regime is not satisfactory. We will propose in this paper some observations that may shed some light on the roots of these difficulties.

\item in the machine learning context one often faces variants of the following situation~\cite{Hastie_book}: from $n$ examples of an input-output relation $(x^{(i)},z^{(i)})$ with $i \in [n] \equiv\{1,\dots,n\}$ one would like to generalize and predict the output $z^{({\rm new})}$ corresponding to an input $x^{({\rm new})}$ not already seen. In a classical example of such supervised learning problems the inputs are high-dimensional objects, i.e. $x \in \R^d$ with $d \gg 1$, which could be the vectorial representation of a high-resolution image, and the output is low-dimensional, e.g. $z \in \R$ being a label characterizing the type of image in a classification task. Assuming the existence of a function $f$ such that $z^{(i)}=f(x^{(i)},W)$, with $W$ some unknown parameters, and of prior distributions on the $x$'s and $W$, this generalization problem falls under the Bayesian inference setting of a correlated pair $(S,Y)$ of random variables, the role of the signal being played by $S=z^{({\rm new})}$, and of the observations by $Y=(x^{(\rm new)},\{(x^{(i)},z^{(i)})\}_{i \in [n]})$.

\end{itemize}

\subsection{Optimal estimators}
\label{sec_optimal}

We shall return now to the generic setting of a signal-observation pair of random variables $(S,Y)$, and discuss the constructions of estimators $\hS$, which should be functions from the observation space to the signal space, such that $\hS(Y)$ is ``as close as possible'' to the true signal $S$. Of course to give a precise meaning to this notion we need to introduce a distance $d(S,\hS)$ on the signal space that measures the closeness of the signal to its estimation (we will not require $d$ to be a distance in the mathematical sense, namely we only ask that $d$ is non-negative and that $d(S,\hS)=d(\hS,S)$, without imposing the definiteness and triangular inequality for $d$). The Bayesian risk of an estimator is defined as its average distance to the groundtruth signal,
\beq
R(\hS)=\E[ d(S,\hS(Y))] \ ,
\label{eq_def_R}
\eeq
the average being over the joint law of $(S,Y)$ discussed above. A Bayes-optimal (BO) estimator, for a given choice of the distance $d$, is then one which minimizes the risk, $\hS^{\rm BO} \in \text{arginf} \ R(\hS)$. Decomposing the expression (\ref{eq_def_R}) of the risk as an average over $Y$ and a conditional average over $S|Y$ (with the law described in Eq.~(\ref{eq_posterior})), one can perform this minimization for each point of the observation space independently, and obtain formally the expression of the optimal estimator as
\beq
\hS^{\rm BO} (Y) \in \underset{x}{\text{arginf}} \ \E[d(S,x) | Y ] \ ,
\label{eq_hSBO}
\eeq
with $x$ running on the signal space. We insist on the fact, that will play an important role in the following, that the notion of optimality depends on the choice of $d$, i.e. on the way one measures the accuracy of estimation. Two special cases are worth spelling out more explicitly:
\begin{itemize}
\item suppose that the signals are vectors of discrete coordinates, i.e. $S \in \chi^N$ with $\chi$ finite (in the SBM model one would have $\chi = \{1,\dots,q\}$). Then a natural choice for $d$ is the Hamming distance, $d(S,\hS) = \sum_{i=1}^N \one(S_i \neq\hS_i)$, that counts the number of indices for which $\hS$ gives a wrong answer. With this choice one can easily find from (\ref{eq_hSBO}) that the optimal estimator is, for each coordinate $i\in[N]$, $\hS_i^{\rm BO} (Y) \in \underset{x \in \chi}{\text{argmax}} \ \mathbb{P}(S_i=x |Y)$. This is called the symbol Maximal A Posteriori (MAP) estimator in the coding theory literature, since it selects the value with the maximum marginal probability in the posterior measure.

\item if the signals are continuous, $S \in \mathbb{R}^N$, a common choice for $d$ is the square of the $L^2$ norm, that we shall call the Square Error (SE) distance,
\beq
d(S,\hS)= ||S - \hS||_2^2 = \sum_{i=1}^N (S_i - \hS_i)^2 \ , 
\label{eq_dSE}
\eeq
which gives the Mean Square Error (MSE) expression for the risk, $R(\hS) = \E[||S - \hS(Y)||_2^2 ]$. The optimal estimator is then well-known to be the posterior mean $\hS^{\rm BO} (Y) = \E[S|Y]$; this result is easily derived from (\ref{eq_hSBO}) by noting that the optimization can be performed independently for each $i \in [N]$, and that the latter can be done explicitly:
\beq
\hS_i^{\rm BO} (Y) \in \underset{x \in \R}{\text{arginf}} \ \E[(S_i-x)^2 | Y ] \ , \qquad \E[(S_i-x)^2 | Y ]= \E[S_i^2|Y] - 2 x \E[S_i|Y] + x^2 \ ,
\eeq
the minimum of this quadratic function on $x$ being easily found to be reached in $x = \E[S_i|Y]$.
\end{itemize}

In the continuous case this choice of $d$ can be viewed as rather arbitrary: one could take more generically $d(S,\hS)= \sum_i |S_i - \hS_i|^p$, with no compelling reason to favour $p=2$ with respect to other values of $p$. The very important advantage of the choice $p=2$ is however the simple analytic expression it yields for the corresponding optimal estimator, namely $\hS^{\rm BO} (Y) = \E[S|Y]$. For generic values of $p$ the optimal estimator can be seen as a Fr\'echet mean of the posterior distribution, but it does not admit an explicit formula in general (for $p=1$ the conditional median replaces the conditional mean, for generic $p$ \cite{BaDyLiPo24} discusses specific priors such that the optimal estimator is linear in $Y$). We will come back later on this point, arguing that symmetries may force one to abandon the usual square error distance for $d$, and as a consequence loose the simplicity of the corresponding optimal estimator.

\subsection{Symmetries}

The notion of symmetry is formalized mathematically in terms of a group, namely a set $G$ endowed with an associative multiplication rule admitting a neutral element $e$ with $eg =ge =g$ for all $g \in G$, and such that each element $g$ admits an inverse $g^{-1}$ with $gg^{-1}=g^{-1} g =e$. A group $G$ is said to act on a set $U$ if there exists an application that associates to each $g \in G$ and $u \in U$ an element of $U$ denoted $g \cdot u$, interpreted as the image of $u$ under the transformation $g$, that respects the multiplication rule of the group, in the sense that $g\cdot (h \cdot u) = (gh) \cdot u$, and such that the neutral element of $G$ leaves $U$ unchanged, $e \cdot u =u$. Linear representations are an important special case of such group actions, when $U$ is a vector space and the group elements $g$ acts on $U$ by application of linear operators $\rho(g)$. A trivial action can always be defined for any group on any set, by  $g \cdot u =u$. One defines the orbit of an element $u$ under the action of a group $G$ by $G\cdot u = \{g \cdot u \ : \ g \in G\}$, that contains all possible images of $u$ under the transformations of $G$. A function $f:U\to V$, where $U$ is endowed with an action of $G$, is said to be invariant if $f(g \cdot u) = f(u)$ for all $g \in G$ and $u \in U$. If in addition the set $V$ also admits an action of the same group $G$ (that we shall denote with the same symbol $\cdot$ even if in general the underlying applications are distinct), one says that the function $f$ is equivariant (or covariant) if $f(g\cdot u)=g \cdot f(u)$. The definition of invariant functions is thus a special case of the equivariant ones, when the action of $G$ is trivial on $V$. The characterization of invariant and equivariant functions, and most notably polynomials, is an important branch of mathematics known as invariant theory~\cite{GoWa_book}, with many results in particular when $G$ is the symmetric group of permutations acting on the coordinates of a function of several variables~\cite{McDo98}, or other classical groups like the orthogonal one~\cite{We66}.

\subsection{Symmetries of inference problems}

Putting together the two ingredients introduced above, we shall say that a group $G$ is a symmetry of an inference problem if it acts on both the signal and observations spaces in such a way that $(g \cdot S,g\cdot Y) \eqd (S,Y)$ for all $g \in G$, with $\eqd$ denoting equality in distribution of random variables. Let us insist on the facts that in this definition the same $g$ acts simultaneously on the signal and on the observations, and that even if one uses the same symbol $\cdot$ the two actions are in general distinct (even the two spaces of signal and observations are generically different). In terms of the two laws $P_S$ and $P_{Y|S}$ we used above to define the random variables, this definition amounts to $P_S(g \cdot s)=P_S(s)$ and $P_{Y|S}(g \cdot y | g \cdot s) = P_{Y|S}(y|s)$ for all $g \in G$.

In the next two Sections we shall explore the consequences of such symmetries in inference problems, arguing that they can be helpful in some cases, or detrimental in others. In colloquial terms, the fuzzy boundary between these two sides depends on the ``relative strength'' of the actions on $G$ on the signal and observations spaces: as an oversimplified rule of thumb, the more $G$ acts on the observations (i.e. the bigger the orbits it induces in this space), the better, while on the contrary the more $G$ acts on the signal, the worst. We will give more flesh to these imprecise statements in the rest of the manuscript, at this point an intuitive justification can be provided as follows: the action of $G$ on $S$ can be interpreted as a blurring of the sought-for signal, hence as a loss of information, if it does not come along with an action on the observations $Y$, that are the only directly accessible sources of knowledge on $S$.

\section{Symmetries as an asset}
\label{sec_asset}

In all this Section we assume that the signal is continuous, $S \in \R^N$ (even if in some cases $S$ can have an internal structure, like matrix indices, it can always be seen as a vector), and that the quality of estimation is measured in terms of the Square Error distance $d$ defined in Eq.~(\ref{eq_dSE}). We further assume that a group $G$ is a symmetry of the inference problem, in the sense defined above, and that its action on the signal space is a linear isometric representation (i.e. $g\cdot S = \rho(g) S$, with $\rho(G)$ an $N \times N$ orthogonal matrix).

As discussed previously this choice of $d$ implies that the Bayes-optimal estimator is the posterior mean, $\hS^{\rm BO} (Y) = \E[S|Y]$. It is a simple exercise to check that the symmetry hypothesis leads to the equivariance of this optimal estimator, i.e. $\hS^{\rm BO} (g \cdot Y) = g \cdot \hS^{\rm BO} (Y) $. If the problem is sufficiently simple for the conditional mean to be computable exactly (or at least asymptotically exactly, for instance in a large dimension limit), this equivariance property is merely a sanity check of the correctness of the result. There are however problems where a direct computation of $\hS^{\rm BO}$ is challenging; we shall describe now an alternative approach, known as the low-degree polynomial strategy~\cite{ScWe22,MoWe22,Se24}, that can be useful in such cases, and where the symmetries play a very helpful role.

The idea of this method is to look for an approximation of $\hS^{\rm BO}$ in a variational way, considering the estimators of the form
\beq
\hS(Y) = \sum_{\beta \in \A} c_\beta b_\beta(Y) \ ,
\label{eq_CL}
\eeq
where the $b_\beta$ are simple functions from the observation space to $\R^N$, $\A$ is a finite index set, and the $c_\beta$ are free parameters in this linear combination. The latter are chosen to minimize the risk of $\hS$, which corresponds here to the MSE $R(\hS) = \E[||S - \hS(Y)||_2^2 ]$. Inserting the variational expression (\ref{eq_CL}) one obtains $R(\hS) = \E[||S||_2^2] + c^T \M c -2 \cR^T c$, where $c=\{c_\beta\}_{\beta \in \A}$ and $\cR$ are vectors of size $|\A|$, $\M$ is a square matrix of the same size, the elements of $\cR$ and $\M$ being given by
\beq
\M_{\beta,\beta'} = \E[\la b_\beta(Y),b_{\beta'}(Y) \ra ] \ , \qquad
\cR_\beta = \E[\la S,b_\beta(Y) \ra] \ ,
\eeq
where $\la S,T \ra = \sum_{i=1}^N S_i T_i$ is the canonical scalar product in $\R^N$. The optimal estimator, within this restricted variational space of linear combinations of trial functions indexed by $\A$, is thus obtained by minimizing the quadratic function of the free parameters $c$, yielding the linear system of equations $\M c=\cR$. In general it is not too difficult to compute the elements of $\M$ and $\cR$, the complexity of this approach can thus be much lower than the direct computation of the conditional mean. In practice the determination of the elements of $\M$ and $\cR$ is reduced to the computation of (joint) moments of $S$ and $Y$ when the basic functions $b_\beta$ of the decomposition are taken as polynomials in the observations (assuming $Y$ is a vector of real random variables). This explains the name ``low-degree polynomial'' of the method (which initially emerged in the context of hypothesis testing problems, see~\cite{HoSt17,KuWeBa19}).

It should be clear that the quality of the approximate estimator $\hS$ obtained in this way improves (i.e. that its MSE decreases) if one includes more and more functions in the trial set $\{b_\beta\}_{\beta \in \A}$; at the same time the complexity of the computation, and in particular the size of the linear system of equations $\M c=\cR$, also increases along the way, requiring to settle for a compromise between these two considerations. It turns out that symmetries have a very beneficial consequence here: one does not pay any price in terms of quality, and may gain a lot in terms of complexity, by focusing on equivariant basic functions $b_\beta$. This statement is quite intuitive: since our objective was to emulate $\hS^{\rm BO}$, which is itself equivariant, it is not so surprising that its best approximation is generated by linear combinations of equivariant functions. One can give a formal proof of this fact (see Appendix B.3 in~\cite{MoWe22}, where it is called the Hunt-Stein lemma, and Sec. III.C of~\cite{Se24}), by first showing that for any estimator the risk cannot increase under symmetrization: $R(\hS^{\rm eq}) \le R(\hS)$, where $\hS^{\rm eq}$ is the equivariant function defined by
\beq
\hS^{\rm eq}(Y) =\E_g[g^{-1} \cdot \hS(g \cdot Y) ] \ ,
\label{eq_projection}
\eeq
the expectation being on the invariant (Haar) measure on $G$. The mapping $\hS \to \hS^{\rm eq}$ defined in (\ref{eq_projection}) is a linear transformation in the space of functions from observations to signals, that projects onto the subspace of equivariant functions. Since $R(\hS^{\rm eq}) \le R(\hS)$ the MSE reached by a variational family $\{b_\beta\}_{\beta \in \A}$ cannot be strictly better than the one of $\{b_\beta^{\rm eq}\}_{\beta \in \A}$; the latter can nevertheless be of a strictly smaller cardinality, if originally distinct $b_\beta$ are projected to the same element by the symmetrization (\ref{eq_projection}), hence a reduction of the computational cost with no loss in estimation accuracy.

Let us give a few examples of this phenomenon:
\begin{itemize}

\item The first one is almost trivial but has some pedagogical virtue. Consider indeed a scalar denoising problem: both signal and observation are real random variables, and the observation channel is $Y=S+Z \in \R$ with the noise $Z$ independent of $S$. The Bayes-optimal estimator $\hS^{\rm BO}(Y)=\E[S|Y]$ has no closed form expression in general (an exception arising when $S$ and $Z$ are both Gaussian), one can thus try a polynomial approximation of degree $D$ with $\hS(Y) = c_0 + c_1 Y + \dots + c_D Y^D$. According to the generic formulation above the optimal value of the coefficients $c$ will be obtained from the solution of $\M c=\cR$, where $\M$ is a $(D+1) \times (D+1)$ matrix with elements $\M_{p,p'} = \E[Y^{p+p'}]$, and $\cR$ the vector with $\cR_p=\E[S Y^p]$. Suppose now that the signal and noise have even distributions, in the sense that $S \eqd -S$, $Z \eqd -Z$. As a consequence $(S,Y) \eqd (-S,-Y)$, which shows that the multiplicative group $G=\{-1,1\}$, acting on the signal and observation space by multiplication, is a symmetry of the inference problem. The equivariant functions in this toy model are odd in the usual sense (they are such that $f(-Y)=-f(Y)$), the symmetry considerations allows us to reduce a priori the variational functions to the odd monomials and work with $\hS(Y) = c_1 Y + c_3 Y^3 + \dots$; equivalently, the even monomials vanish under the symmetrization by Eq.~(\ref{eq_projection}). In other words the use of the symmetry divides the number of the trial functions by 2, for the same maximum degree $D$, and hence the same quality of approximation. Of course in this trivial case one could have considered all polynomials, and discovered very easily that $c_2=c_4=\dots=0$ in the minimizers, but in more complicated situations the identification of the relevant terms is not so obvious, and the gain in the size of the trial set can be much more substantial.

\item Such non-trivial simplifying aspects of symmetries have been exploited in the low-rank matrix factorization problem in~\cite{MoWe22}, and in the matrix denoising one in~\cite{Se24}. Recall that in the latter case one has also $Y=S+Z$, but now these objects are $n \times n$ matrices, and the objective is to send the size $n$ to infinity. In this context a polynomial estimator $\hS(Y)$ is a matrix whose elements $\hS(Y)_{i,j}$ are multivariate polynomials in the entries of $Y$, i.e. in $Y_{1,1}, Y_{1,2} , \dots , Y_{n,n}$. One thus faces the difficulty that, even for a fixed maximal degree $D$, the number of such monomials increases with $n$, and as a consequence the size $|\A|$ of the matrix $\M$ and vector $\cR$ of the variational approach diverges in the large $n$ limit, making the whole procedure quite impractical. Fortunately symmetry considerations allow to escape this deadend: the probability laws of several classical ensembles of RMT are invariant under the conjugation by the orthogonal group (i.e. under $M \to O M O^T$ for all $O$ with $OO^T=O^TO=\one$). If $S$ and $Z$ enjoy this invariance in law then the orthogonal group becomes a symmetry of the inference problem. One can then restrict the polynomial estimators $\hS(Y)$ to the equivariant ones, without loosing on the accuracy according to the generic discussion above. This is actually a strong requirement (in colloquial terms, the action of this group induces ``big enough'' orbits) and the number of equivariant polynomials is drastically reduced with respect to the number of arbitrary ones. One can indeed show that the equivariant polynomials of degree at most $D$ are linear combinations of
\beq
Y^p (\Tr \, Y)^{q_1} (\Tr(Y^2))^{q_2} \dots (\Tr(Y^D))^{q_D} \ ,
\label{eq_On_equiv}
\eeq
for non-negative integers $p,\{q_i\}$, the total degree of the polynomial being $p+\sum_i i q_i \le D$. Note that the dimension of this subspace of equivariant polynomials depends on $D$ but not on $n$, which allows to perform the large size limit in the variational approach with $\M$ and $\cR$ of fixed sizes. The hypothesis of invariance under orthogonal conjugation is violated for some ensembles of random matrices, like the arbitrary Wigner ones with i.i.d. non-Gaussian matrix elements, that however preserves a weaker subgroup of the orthogonal one, namely their laws are invariant by permutation of the row and column indices. The equivariant polynomials under the permutation group are more numerous than those under the orthogonal group written in (\ref{eq_On_equiv}), but their number is still independent of $n$ for a fixed degree, and they can be indexed by suitable finite graphs; see~\cite{MoWe22,Se24} for more details on this point, and~\cite{KuMoWe24} for an extension to the tensor case.

\item Symmetries can also play a simplifying role in the machine learning setting defined in Sec.~\ref{sec_def}, where $S=z^{({\rm new})}$ is to be inferred from $Y=(x^{(\rm new)},\{(x^{(i)},z^{(i)})\}_{i \in [n]})$. In this context it is indeed natural to assume that the function $z=f(x)$ is invariant under the action of a group $G$ on the inputs $x$: for image classification the label $z$ should not change if the input image $x$ is translated, or slightly deformed. In the language of the present paper this means that $G$ is a symmetry of the inference problem with a trivial action on the signal and a non-trivial action on the observations. This is somehow the ideal situation, in the sense that the reduction of the number of candidate variational functions is the strongest possible and collapse to a projection on the invariant functions. There is in addition a permutation symmetry on the indices $i$ of the sample pairs $\{(x^{(i)},z^{(i)})\}_{i \in [n]}$ that can be further exploited. This phenomenon was studied and put on a quantitative basis in~\cite{MeMiMo21}, through the definition of the \emph{degeneracy} of a group in terms of the asymptotic reduction in the number of invariant polynomials compared to arbitrary ones. To be precise the study of~\cite{MeMiMo21} does not treat the Bayes-optimal estimation, but shows the reduction of sample complexity or network width induced by the degeneracy for some specific architectures (random features and ridge kernel regression).

\end{itemize}

%\bigskip

%{\bf make the link with the intuitive interpretation with the ``sizes'' of the actions, how much does the symmetrization (\ref{eq_projection}) reduces the possible variational set ? Limit cases with trivial action on the signal vs observation.}

\section{Symmetries as a nuisance}
\label{sec_nuisance}

\begin{figure}
\begin{tikzpicture}
\draw[-latex] (-2.4,0) -- (3,0) node [below right] {$S_1$};
\draw[-latex] (0,-2.2) -- (0,3) node [above right] {$S_2$};
\fill[black] (2.5,2.5) circle (2pt) node [right] {$\hS$};
\fill[black] (-1.5,1) circle (2pt) node [left] {$S$};
\draw (0,0) circle [radius=1.802775]; 
\draw[dashed,latex-latex] (-1.5,1) -- (2.5,2.5) node [near end,above,sloped] {$d$};
\draw[dashed,latex-latex] (1.27475487,1.27475487) -- (2.5,2.5) node [midway,below] {$d_G$};
\end{tikzpicture}
\caption{A sketch illustrating the definition of the quotiented distance of Eq.~(\ref{eq_quotient}).}
\label{fig_quotient}
\end{figure}
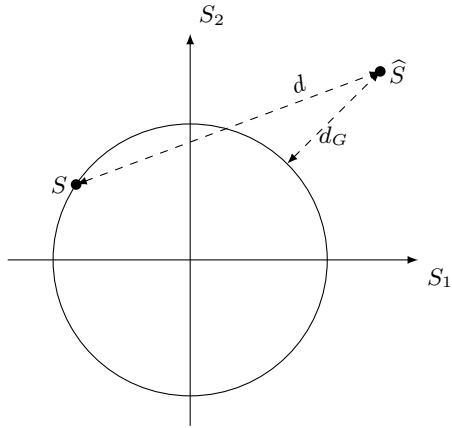

We will now describe somehow opposite situations, in which the presence of a symmetry induces difficulties in the inference problems. According to the informal discussion above this is to be expected when the action of $G$ is stronger on the signal than on the observation. We shall in fact make in this section the drastic assumption that $G$ acts trivially on the observations, i.e. that $(g\cdot S,Y) \eqd (S,Y)$ for all $g \in G$, with a non-trivial action on $S$. In terms of the laws defining the process this means that the prior is invariant, $P_S(g \cdot s)=P_S(s)$, and that $P_{Y|S}(y|g \cdot s) = P_{Y|S}(y|s)$. It should be clear that the ``signal'' that can be recovered in such a situation is not $S$ itself, but rather the orbit $G \cdot S = \{g \cdot S \ : \ g \in G \}$, since all the points of this orbit generate observations with the same law, and are thus perfectly indistinguishable from the knowledge available to the observer. An example that is visually easy to grasp is provided in Fig.~\ref{fig_quotient}: the signal space is here $S=(S_1,S_2) \in \mathbb{R}^2$, and $G={\rm SO}(2)$ acts by rotating this plane, in such a way that the orbits are circles centered on the origin.

We insisted in Sec.~\ref{sec_optimal} on the crucial role played by the choice of the distance $d(S,\hS)$, and in particular on the dependency of the optimal estimator on $d$. This point is particularly relevant here: if for instance one uses the SE distance in the problem sketched in Fig.~\ref{fig_quotient} one obtains as an optimal estimator $\hS^{\rm BO}(Y)=\E[S|Y]=0$, since the posterior is uniform on the circles and then averages out to the origin of the plane. This apparent contradiction that the ``optimal'' estimator is a trivial constant even if $Y$ contains a non-trivial information arises from the incorrect choice of the distance $d$ used. In this very simple example the way out is easy to find: one can use polar coordinates $(r,\theta)$ in the plane instead of the cartesian ones $(S_1,S_2)$, notice that the orbits are in bijection with the radius $r$, hence take for $d(S,\hS)$ any distance between $r$ and $\hr$, independently of $\theta$ and $\htheta$. However for high-dimensional signal spaces, and complicated group actions on them, it is not obvious to find an explicit parametrization of the space that allows to separate the coordinates of the orbit from the coordinates inside an orbit. In this case a generic prescription is to start from a distance $d$ (possibly too fine, i.e. that separate points of the same orbit), and to quotient it by the action of the group, defining a new distance $d_G$ with
\beq
d_G(S,\hS) = \inf_{g \in G} d(g\cdot S , \hS) \ .
\label{eq_quotient}
\eeq
We assume that $d$ and the action of $G$ are such that $d(g\cdot S , \hS)=d( S , g^{-1} \cdot \hS)$, which yields the symmetry $d_G(S,\hS)=d_G(\hS,S)$, and the independence of $d_G$ on the orbit representant, namely $d_G(g \cdot S,\hS)=d_G(S,\hS)$. In the example of Fig.~\ref{fig_quotient} the quotient of the SE distance $d(S,\hS)=(S_1-\hS_1)^2 + (S_2-\hS_2)^2$ by the rotation group according to Eq.~(\ref{eq_quotient}) yields $d_G(S,\hS)=(r-\hr)^2$, a valid distance between orbits in agreement with the intuitive reasoning above.

We emphasized in Sec.~\ref{sec_optimal} that very few choices of the distance $d$ yield a closed-form expression for the corresponding Bayes-optimal estimator $\hS^{\rm BO}$. As a consequence the presence of a symmetry, and hence the need to trade the distance $d$ by its coarse-grained version $d_G$, will in general complicate the resolution of such inference problems (even when the initial distance measure was the SE or Hamming ones). We will present later on a series of explicit examples, but before that we point out some previous works where this phenomenon was already discussed.

A well-known case where this kind of symmetry occurs is the SBM introduced in Sec.~\ref{sec_def_inference}, in particular in its symmetric version where the prior distribution of the labels is uniform, $\oeta=(\frac{1}{q},\dots,\frac{1}{q})$, and where the elements of the affinity matrix $c_{\s,\tau}$ only depend on whether $\s=\tau$ or not (i.e. $c$ is constant on the diagonal, and off-diagonal). In that case it is easy to check that $G={\cal S}_q$, the symmetric group of permutations of $[q]$, acting on the labels $S=(\s_1,\dots,\s_N)$ by $\pi \cdot S = (\pi(\s_1),\dots,\pi(\s_N))$, and leaving the observed graph untouched, is a symmetry of the problem. This was of course noticed in~\cite{Decelle_SBM_long}, which used as a measure of the distance between the signal and its estimation the quotiented version of the Hamming distance, according to the definition (\ref{eq_quotient}) with $G={\cal S}_q$. A non-trivial estimation, in the large size limit, is obtained if this (normalized) quotiented distance is strictly smaller that what can be achieved by a dummy estimator that does not access $Y$, and only uses the prior information on $S$. There has been some discussions, see in particular~\cite{Abbe_review_SBM}, on the appropriate generalization of this definition when the parameters $\oeta$ and $c$ are only partially symmetric. In~\cite{Abbe_review_SBM} it was emphasized that the definition (\ref{eq_quotient}) with $G={\cal S}_q$ was not satisfactory in this case, and proposed an alternative definition based on the possibility to partition the vertices in two sets which have a different empirical content of at least two communities. We believe that (\ref{eq_quotient}) could also be used in the partially symmetric case, provided one takes in this definition $G$ as the largest subgroup of ${\cal S}_q$ with $(g\cdot S,Y) \eqd (S,Y)$ for all $g \in G$ (which will be a proper subgroup of ${\cal S}_q$ if there is only a partial symmetry in the parameters of the SBM). In the statistical mechanics treatment of~\cite{Decelle_SBM_long} the symmetry difficulty was bypassed thanks to the spontaneous breaking of the posterior measure into various ``pure states'' related one to the other by the action of $G$, i.e. by the permutations of the labels; in technical terms the cavity (or Belief Propagation) equations admit a trivial symmetric solution, that would yield the Bayes optimal estimator if the naive Hamming distance were to be used, and several symmetry breaking solutions (corresponding to each pure states). Picking one of the latter allows to solve the Bayes-optimal estimation with the quotiented distance (\ref{eq_quotient}). Notice that here the size of the symmetry group is at most $q!$, and thus remains finite in the large graph limit (we will see later on an example where this will not be the case).

We would also like to mention here the research field called orbit recovery, or estimation under a group action, see for instance~\cite{AbPeSi18,BaBlKiNiPeWe23} and reference therein, that is tightly related to the present discussion (see also~\cite{BuDoLeMaWaWi24} for computational aspects of related problems). In this setting one generates an observation $Y=(Y_1,\dots,Y_n)$ through
\beq
Y_i = g_i \cdot S + \xi_i \ ,
\eeq
with $g_i$ i.i.d. copies of an uniformly chosen member of a group $G$ (i.e. the $g_i$ are drawn from the Haar measure on $G$), and $\xi_i$ some i.i.d. noise variables, with intensity controlled by a parameter $\Delta$. The motivation for these studies comes from applications to cryo-electron microscopy (cryoEM)~\cite{AdDuLeMc84,SiSh11}, an experimental method where successive projections of randomly rotated molecules are observed, or its abstraction called multi-reference alignment (MRA)~\cite{Sig98}. As discussed above the signal $S$ is here indistinguishable from any member of the orbit $G \cdot S$, since the $g_i$'s are uniform on $G$. There is however an important difference in the perspective taken by these works, which concentrate on the dependence of the sample complexity on the noise intensity $\Delta$, i.e. on how large $n$ should be when $\Delta$ diverges in order to recover the orbit of $S$ within a fixed accuracy. On the contrary here we consider a single sample, and focus on the Bayes-optimal estimation one can extract from it.

The procedure to achieve this task is in principle clear: inserting the definition (\ref{eq_quotient}) of the quotiented distance in the expression (\ref{eq_hSBO}) of the Bayes-optimal estimator yields
\beq
\hS^{\rm BO} (Y) \in \underset{x}{\text{arginf}} \ \E\left[ \ \inf_{g \in G} d(g\cdot S , x) \ \Big| \ Y \right] \ ;
\label{eq_hSBO_dG}
\eeq
note that by definition this optimal estimator is not unique, if $\hS(Y)$ minimizes the average quotiented distance (for a given $Y$) this will also be the case of all members of the orbit $G \cdot \hS(Y)$. The actual computation of (\ref{eq_hSBO_dG}) seems daunting in general, as it relies on two extremizations (over $g$ and $x$) intertwined with an average (over $S|Y$). We shall now consider a few concrete examples where one can give more explicit expressions of $d_G$ and of the corresponding $\hS^{\rm BO}$.

\begin{itemize}

\item Our first example will be elementary; we consider a scalar signal $S \in \R$, and the SE distance $d(S,\hS)=(S-\hS)^2$ as the naive measure for the accuracy of estimation. The multiplicative group $G=\{-1,1\}$, acting by multiplication on $S$ (i.e. $g \cdot S = g S$ for $g=\pm 1$), induces the quotiented distance
\beq
d_G(S,\hS)=\min((S-\hS)^2,(S+\hS)^2) = S^2 + \hS^2 - 2 \max(S \hS,-S \hS) = S^2 + \hS^2 - 2 |S| |\hS| = (|S| - |\hS|)^2 \ ,
\eeq
which one recognizes as a valid distance between the orbits $G \cdot S = \{|S|,-|S| \}$ and $G \cdot \hS$. To obtain the Bayes-optimal estimators for this quotiented distance we compute
\beq
\E[(|x|-|S|)^2 | Y] = x^2 - 2 |x| \E[|S| | Y] + \E[S^2 | Y] \ ,
\eeq
which is minimized in $x = \E[|S| | Y]$ and $x = - \E[|S| | Y]$. This yields the two optimal estimators,
\beq
\hS^{\rm BO}(Y) = \E[|S| | Y] \quad \text{and} \ \ \hS^{\rm BO}(Y) = - \E[|S| | Y] \ ,
\eeq
as could be expected intuitively. As an example of a pair $(S,Y)$ admitting this symmetry one can take for $S$ a symmetric real random variable (i.e. with $S \eqd - S$), and observe it through a multiplicative noise with symmetric distribution, namely defining the observation as $Y =Z S$ with $Z$ independent of $S$, and such that $Z \eqd -Z$. As in the example of Fig.~\ref{fig_quotient} the Bayes-optimal estimator for the naive SE distance $d$ is trivial, $\E[S|Y]=0$, even when $Y$ contains a lot of information on the orbit of $S$ (and potentially allows to determine it exactly, as for example when $Z = \pm 1$ with equal probabilities $1/2$).

\item The example above was very simple, and led to an optimal estimator which could be expressed in terms of a posterior mean, not of the signal itself, but of a simple function of $S$ (here the absolute value). It is however instructive as it shows that such a situation should be considered as a fortunate coincidence. Consider indeed the following slight variation of the previous example: now $S \in \R^N$ with $N \ge 2$, with $d(S,\hS)=||S-\hS||_2^2$ and again $G=\{-1,1\}$ acting multiplicatively on $S$. The quotiented distance thus reads
\beq
d_G(S,\hS)=\min(||S-\hS||_2^2,||S+\hS||_2^2) = ||S||_2^2 + ||\hS||_2^2 - 2 | \la S, \hS \ra | \ .
\eeq
At variance with the $N=1$ case the absolute value of the scalar product $S_1 \hS_1 + \dots + S_N \hS_N$ cannot be factored as the product of a function of $S$ and one of $\hS$. As a consequence in the expression of the optimal estimator
\beq
\hS^{\rm BO} (Y) \in \underset{x \in \R^N}{\text{arginf}} \ \left( ||x||_2^2 - 2 \, \E\left[ \ | \la S,x \ra | \ \Big| \ Y \right] \right) \ ,
\eeq
the minimization on $x$ cannot be performed explicitly (for a generic posterior law of $S|Y$), and $\hS^{\rm BO} (Y)$ is not expressible as the posterior mean of some simple function of $S$.

\item We will now present another fortunate case in which analytical simplifications are possible even though the action of the group is much less trivial than above. We consider vectorial signals, $S \in \R^N$, with the usual SE distance $d(S,\hS)=||S-\hS||_2^2$, and assume now that the symmetry group of the inference problem is $G={\cal S}_N$, the group of permutations of $[N]$, acting on vectors by permutation of the coordinates: for $\pi \in {\cal S}_N$, $(\pi \cdot S)_i = S_{\pi^{-1}(i)}$. This is indeed an action, isometric with respect to $d$ in the sense that $d(\pi \cdot S,\hS)= d(S,\pi^{-1} \cdot \hS)$. We consider the quotiented distance $d_G$ from Eq.~(\ref{eq_quotient}), that reads
\begin{align}
  d_G(S,\hS) & = \min_{\pi \in {\cal S}_N } \sum_{i=1}^N[(S_{\pi^{-1}(i)})^2 + (\hS_i)^2 - 2 S_{\pi^{-1}(i)}\hS_i  ] \\
  & = ||S||_2^2 + ||\hS||_2^2 - 2 \max_{\pi \in {\cal S}_N }\sum_{i=1}^N S_i \hS_{\pi(i)} \ .  \label{eq_d_rearr}
\end{align}
It turns out that the last maximization can be performed rather explicitly, thanks to the so-called rearrangement inequality, see for instance theorem 368 in~\cite{HaLiPo34}. To state this result let us denote $S_{[1]} \le S_{[2]} \le \dots \le S_{[N]}$ a permutation of $S_1,\dots,S_N$ that puts the coordinates in non-decreasing order, and $S^\uparrow$ the corresponding vector, with $(S^\uparrow)_i =S_{[i]}$. The same notation will be used for $\hS$ as well. The rearrangement inequality states that the maximum in the last term of (\ref{eq_d_rearr}) is reached when $\pi$ orders the elements of $\hS$ in the same way as the one of $S$. More explicitly, assuming without loss of generality that $S_1 \le \dots \le S_N$, the maximum is reached for $\pi$ such that $\hS_{\pi(i)} = \hS_{[i]}$. We can thus write
\beq
d_G(S,\hS) = \sum_{i=1}^N [ (S_i)^2 + (\hS_i)^2 - 2 S_{[i]} \hS_{[i]} ] = d(S^\uparrow,\hS^\uparrow) \ ,
\eeq
in other words the quotiented distance coincides with the usual SE distance on the ordered version of the vectors.

As a consequence the Bayes-optimal estimator for this $d_G$ can be computed explicitly, $\hS^{\rm BO}(Y) = \E[S^\uparrow|Y]$, or in components $(\hS^{\rm BO}(Y))_i = \E[S_{[i]}|Y]$ (of course any permutation of this estimator achieves the same risk and is also optimal). This provides another example, less trivial than the one above, of a quotiented distance whose optimal estimator can be expressed as a posterior mean of some function of the signal (here the order statistics of $S$). The practical implementation of such a formula might still be complicated, in particular in the large $N$ limit, but at least one has an explicit expression to start with.

To complete the discussion of this example let us now define a pair of random variable $(S,Y)$ that fulfils this symmetry assumption. We shall take $S_1,\dots,S_N$ as independent identically distributed copies of an absolutely continuous random variable (hence with probability 1 the $S_i$ are distinct, which simplifies the discussion below). The law of $S$ is clearly invariant under ${\cal S}_N$. Consider in addition the observation channel $Y=(Y_1,\dots,Y_N)$, where for $p \in [N]$:
\beq
Y_p = \sum_{i=1}^N (S_i)^p \ .
\eeq
The $Y_p$'s are deterministic functions of $S$, invariant under the permutations of coordinates of $S$, hence one has indeed $(\pi \cdot S,Y) \eqd (S,Y)$. The theory of symmetric functions~\cite{McDo98} (in which the $Y_p$'s are known as the power sum symmetric polynomials) demonstrates that the knowledge of $Y$ allows to determine the set $\{S_1,\dots,S_N\}$ (but of course not the ordered vector). The posterior law $P(\cdot | Y)$ is thus the linear combination of $N!$ atoms, with equal weights, on the permutations of the vector $S$ corresponding to this realization of $Y$. The risk of the Bayes-optimal estimator for the quotiented distance thus vanishes, since the orbit of $S$ is perfectly reconstructed from $Y$. On the contrary if one had used the naive SE distance the optimal estimator would have been $(\hS^{\rm BO}(Y))_1= \dots = (\hS^{\rm BO}(Y))_N = \frac{1}{N} Y_1$, giving a risk proportional to the variance of the law of the $S_i$'s.

\item The last example of a detrimental symmetry we shall consider now is motivated by the extensive rank matrix factorization problem~\cite{SaKa13,KaKrMeSaZd16,Sc18,MaKrMeZd22,BaMa22,TrErKrMaZd22,PoBaMa23,BaKoRa24,PoMa23,PoMa23b,LaMeGa23,CaMe23a,CaMe23b} we mentioned in Sec.~\ref{sec_def_inference}. We recall that in this setting the signal, denoted $X$ to avoid confusion with the matrix denoising problem, is an $n \times r$ matrix observed through the channel $Y = XX^T +Z$, the observation $Y$ being an $n \times n$ matrix, and $Z$ modelling an additive noise independent of $X$. Denoting $X_{i,\mu}$ the matrix elements of $X$, one has more explicitly
\beq
Y_{i,j} = \sum_{\mu=1}^r X_{i,\mu} X_{j,\mu} + Z_{i,j} \ ,
\eeq
for $i,j \in [n]$. Suppose now that in the prior law on $X$ the elements $X_{i,\mu}$ are i.i.d. random variables (one could actually make the weaker assumption that the vectors $X_\mu = \{X_{i,\mu}\}_{i\in[n]}$ are i.i.d. vectors in $\R^n$); then the permutation group $G = {\cal S}_r$, acting on $X$ as $(\pi \cdot X)_{i,\mu} = X_{i,\pi^{-1}(\mu)}$ for $\pi \in G$, is a symmetry of the inference problem in the sense that $(\pi \cdot X,Y)\eqd (X,Y)$. Since its action is trivial on the observation this is an example of a detrimental symmetry in which the distance measuring the accuracy of estimation has to be amended in order to gauge the non-trivial unobservable orbits of the signal (this symmetry acting on the ``internal'' indices $\mu$ of $X$ should not be confused with the one of the ``external'' indices $i$, that was shown to be beneficial in the matrix denoising problem in Sec.~\ref{sec_asset}).

The natural naive distance is here the squared Frobenius norm of the difference between $X$ and its estimator $\hX$, or in other words the sum of the SE for each matrix element,
\beq
d(X,\hX) = \sum_{i=1}^n \sum_{\mu=1}^r (X_{i,\mu} - \hX_{i,\mu})^2 \ .
\eeq
The optimal estimator with respect to $d$ is as usual the posterior mean $\hX(Y)=\E[X|Y]$; its components $\hX_{i,\mu}(Y)$ are independent on $\mu$ because of the invariance under $G$, which would yield a rather unsatisfactory estimate $\hX(Y)$ of rank 1 (that would actually vanish identically if in addition the law of $X$ is assumed to be even, i.e. if $X \eqd -X$, since $Y$ is invariant under a change of sign in $X$). As explained in generic terms above one should use instead the quotiented version of $d$, according to the definition (\ref{eq_quotient}) for the group $G = {\cal S}_r$, that reads (see also Sec.~V of~\cite{BaCaKoOk24} for a discussion of this distance)
\beq
d_G(X,\hX) = \sum_{i=1}^n \sum_{\mu=1}^r [(X_{i,\mu})^2 + (\hX_{i,\mu})^2] - 2 \max_{\pi \in {\cal S}_r} \ \sum_{i=1}^n \sum_{\mu=1}^r  X_{i,\mu} \hX_{i,\pi(\mu)} \ .
\eeq
The last term is of the form
\beq
\max_{\pi \in {\cal S}_r} \ \sum_{\mu=1}^r T_{\mu,\pi(\mu)} \ , \quad \text{with} \quad T_{\mu,\nu} = \sum_{i=1}^n X_{i,\mu} \hX_{i,\nu} = \la X_\mu , \hX_{\nu} \ra \ ,
\eeq
where in the last equality we used $\la \cdot, \cdot \ra$ to denote the standard scalar product in $\R^n$. One can recognize here a famous optimization problem, known as the bipartite matching, or linear assignment problem. This can be solved in a time polynomial in $r$~\cite{EdKa75} (in~\cite{BaCaKoOk24} a greedy procedure was used to approximate this optimization), but in general does not admit an explicit analytical solution (one cannot use the rearrangement inequality anymore, apart from the trivial case $n=1$). As a consequence the Bayes-optimal estimator for the quotiented distance takes its generic form (\ref{eq_hSBO_dG}), which reads here
\beq
\hX^{\rm BO} (Y) \in \underset{x \in \R^{n \times r}}{\text{arginf}} \ \left\{ \sum_{i=1}^n \sum_{\mu=1}^r (x_{i,\mu})^2 - 2 \, \E\left[ \max_{\pi \in {\cal S}_r} \ \sum_{i=1}^n \sum_{\mu=1}^r  X_{i,\mu} x_{i,\pi(\mu)} \ \Big| \ Y \right] \right\} \ ,
\label{eq_hXBO}
\eeq
with a priori no simplification in terms of the conditional expectation of some explicit function of $X$. As we shall discuss further in the conclusions this seems a particularly challenging problem in the extensive rank case with $r$ proportional to $n$ in the large size limit, that is not expressible in terms of local marginal laws of the posterior probability $P_{X|Y}$.

\end{itemize}

\section{Conclusions}
\label{sec_conclu}

We have collected in this paper a series of simple observations on the effects of invariances in Bayesian inference problems. We have argued that their consequences depend on the relative strength of the actions of the symmetry group on the hidden signal and on the observed quantities. When the latter are sufficiently affected by the symmetry one can exploit the constraints induced on the candidate estimators to reduce the complexity of the low-degree polynomial approach; on the contrary an invariance of the signal without observable effects imposes a revision of the notion of distance between an estimator and the groundtruth, that should be invariant on the orbits of the group.

We believe that this last point may have been overlooked and would deserve deeper investigations, in particular in the context of the statistical mechanics approach to inference problems. As briefly sketched in Sec.~\ref{sec_def_inference} the latter envisions the posterior probability (\ref{eq_posterior}) as a graphical model with dynamical variables in the signal space, the observations corresponding to a planted quenched disorder. The typical properties of such a model, and in particular the quenched free-energy or mutual information $I(S;Y)$, are then obtained via the replica or cavity method, while single-sample quantities  (i.e. for one given realization of the disorder $Y$) like the marginal probabilities of a finite number of variables are accessible by their algorithmic counterparts, namely the Belief Propagation or Approximate Message Passing techniques (we refer the reader to~\cite{NishimoriBook01,Engel_book01,MezardMontanari07,ZdKr16_review,farbeyond} for reviews of these methods). The latter allows to compute Bayes-optimal estimators for the Hamming or SE distance, that are indeed expressible in terms of the local marginals of the posterior, the former giving access to the corresponding average risk (in particular thanks to the I-MMSE relation~\cite{GuShVe05} for Gaussian noise channels). Unfortunately these connections break down when other distances between the signal and its estimation are used, and in particular when a non-trivial group action on the signal space imposes to quotient the distance according to Eq.~(\ref{eq_quotient}): the mutual information $I(S;Y)$ can no longer be related to the average risk, and the optimal estimator is not anymore expressible in terms of local posterior averages.

When the order of the group remains finite in the thermodynamic limit, as is the case in the symmetric SBM, one can rely on the spontaneous breaking of the posterior measure in pure states that shows up in the statistical mechanics treatment as symmetry breaking solutions of the replica or cavity equations and allows to compute (\ref{eq_quotient}) from (\ref{eq_posterior}). It is much less clear that such a phenomenon can occur when the symmetry group grows in the thermodynamic limit, in particular in the case of extensive rank matrix factorization, an observation that could constitute the starting point for future research. We believe that in such a case the computation of Bayes-optimal estimators should instead start from the formula in (\ref{eq_hSBO_dG}) (or (\ref{eq_hXBO}) in the special case of matrix estimation problems invariant under column permutation), and not from the graphical model interpretation of the posterior (\ref{eq_posterior}). This is a much more challenging ``three-level problem'': for a given value of $x$ and $S$ in the signal space one has to solve a minimization problem over $g \in G$, for instance by considering $d(g\cdot S,x)$ as an energy function and introducing a fictitious temperature ultimately sent to 0. The solution of this ``inner problem'' would then provide a function of $(S,x)$, which should be averaged over $S$ according to the posterior law $S|Y$, and finally minimized over $x$. Multi-level problems of a similar kind were studied with statistical mechanics methods in~\cite{AlBrRaZe11,CaZd11}: the inner problem is solved by a message passing algorithm (for the assignment problem of (\ref{eq_hXBO}) this could be done with the BP approach of~\cite{BaShSh08}), these messages constituting the degrees of freedom of the next-level problem. In principle this approach could be followed for the quotiented distance of (\ref{eq_hXBO}) in the extensive rank matrix factorization problem, with the hope that incorporating explicitly the unobservable symmetry in such a way may alleviate the difficulties encountered by the formalism of~\cite{SaKa13,KaKrMeSaZd16}. We leave this specific question, as well as the more generic problem of the possibility of spontaneous breaking of an extensive symmetry group of a posterior measure into pure states, as open issues for future works.

Let us finally underline that even if we concentrated in this work on a static, Bayes-optimal approach to inference problems, symmetries also play a role in the dynamical studies of learning algorithms, see for instance~\cite{MeMoNg18,SaBiCaKrUrZd20,BeGhJa21,GlUr21,HaCh23,DaTrArPeZdKr24} and references therein. In this context they can again be beneficial, allowing in particular to project a dynamical evolution in a high-dimensional space down to a finite number of order parameters, or detrimental whenever the non-trivial recovery of the groundtruth signal requires the breaking of symmetries present in a random, uninformative initial condition.

\bibliography{biblio}

\begin{thebibliography}{10}

\bibitem{FultonHarris_book}
W.~Fulton and J.~Harris.
\newblock {\em Representation theory, a first course}.
\newblock Springer, 1991.

\bibitem{GoWa_book}
R.~Goodman and N.~R. Wallach.
\newblock {\em Symmetry, Representations, and Invariants}.
\newblock Springer, 2009.

\bibitem{Zee_book}
A.~Zee.
\newblock {\em Group theory in a nutshell for physicists}.
\newblock Princeton University Press, 2016.

\bibitem{Toulouse77}
G.~Toulouse.
\newblock Theory of Frustration Effect in Spin-Glasses: {I}.
\newblock {\em Comm. Phys.}, {\bfseries 2}, 115--119 (1977).

\bibitem{To79}
G.~Toulouse.
\newblock Symmetry and topology concepts for spin glasses and other glasses.
\newblock {\em Phys. Rep.}, {\bfseries 49}, 267 (1979).

\bibitem{Ea89}
M.~L. Eaton.
\newblock Group Invariance Applications in Statistics.
\newblock {\em Regional Conference Series in Probability and Statistics},
  {\bfseries 1}, i--133 (1989).

\bibitem{Wi90}
R.~A. Wijsman.
\newblock Invariant Measures on Groups and Their Use in Statistics.
\newblock {\em Lecture Notes-Monograph Series}, {\bfseries 14}, i--218 (1990).

\bibitem{LehmannCasella_book}
E.~L. Lehmann and G.~Casella.
\newblock {\em Theory of Point Estimation, 2d edition}.
\newblock Springer, 1998.

\bibitem{ShTa89}
J.~Shawe-Taylor.
\newblock Building symmetries into feedforward networks.
\newblock In {\em 1989 First IEE International Conference on Artificial Neural
  Networks, (Conf. Publ. No. 313)}, pages 158--162, 1989.

\bibitem{BrBrCoVe21}
M.~M. Bronstein, J.~Bruna, T.~Cohen, and P.~Velickovic.
\newblock Geometric Deep Learning: Grids, Groups, Graphs, Geodesics, and
  Gauges.
\newblock {\em arXiv:2104.13478},  (2021).

\bibitem{ViHoYaKeSc24}
S.~Villar, D.~W. Hogg, W.~Yao, G.~A. Kevrekidis, and B.~Sch{\"o}lkopf.
\newblock Towards fully covariant machine learning.
\newblock {\em Transactions on Machine Learning Research},  (2024).

\bibitem{CoWe16}
T.~Cohen and M.~Welling.
\newblock Group Equivariant Convolutional Networks.
\newblock In M.~F. Balcan and K.~Q. Weinberger, editors, {\em Proceedings of
  The 33rd International Conference on Machine Learning}, volume~48 of {\em
  Proceedings of Machine Learning Research}, pages 2990--2999, New York, New
  York, USA, 20--22 Jun 2016. PMLR.

\bibitem{ChDoLe20}
S.~Chen, E.~Dobriban, and J.~H. Lee.
\newblock A Group-Theoretic Framework for Data Augmentation.
\newblock {\em Journal of Machine Learning Research}, {\bfseries 21}(245),
  1--71 (2020).

\bibitem{SiGeJaSpHoGeBr21}
B.~Simsek, F.~Ged, A.~Jacot, F.~Spadaro, C.~Hongler, W.~Gerstner, and J.~Brea.
\newblock Geometry of the Loss Landscape in Overparameterized Neural Networks:
  Symmetries and Invariances.
\newblock In M.~Meila and T.~Zhang, editors, {\em Proceedings of the 38th
  International Conference on Machine Learning}, volume 139 of {\em Proceedings
  of Machine Learning Research}, pages 9722--9732. PMLR, 18--24 Jul 2021.

\bibitem{NishimoriBook01}
H.~Nishimori.
\newblock {\em Statistical Physics of Spin Glasses and Information Processing:
  An Introduction}.
\newblock Oxford University Press, Oxford, UK, 2001.

\bibitem{Engel_book01}
A.~Engel and C.~Van~den Broeck.
\newblock {\em Statistical Mechanics of Learning}.
\newblock Cambridge University Press, 2001.

\bibitem{MezardMontanari07}
M.~M\'ezard and A.~Montanari.
\newblock {\em Physics, Information, Computation}.
\newblock Oxford University Press, Oxford, 2009.

\bibitem{ZdKr16_review}
L.~Zdeborov\'a and F.~Krzakala.
\newblock Statistical physics of inference: thresholds and algorithms.
\newblock {\em Advances in Physics}, {\bfseries 65}(5), 453--552 (2016).

\bibitem{farbeyond}
P.~Charbonneau, E.~Marinari, M.~M\'ezard, G.~Parisi, F.~Ricci-Tersenghi,
  G.~Sicuro, and F.~Zamponi, editors.
\newblock {\em Spin Glass Theory and Far Beyond}.
\newblock World Scientific, 2023.

\bibitem{fienberg1981categorical}
S.~E. Fienberg and S.~S. Wasserman.
\newblock Categorical data analysis of single sociometric relations.
\newblock {\em Sociological methodology}, {\bfseries 12}, 156--192 (1981).

\bibitem{holland1983stochastic}
P.~W. Holland, K.~B. Laskey, and S.~Leinhardt.
\newblock Stochastic blockmodels: First steps.
\newblock {\em Social networks}, {\bfseries 5}(2), 109--137 (1983).

\bibitem{BoJaRi07}
B.~Bollob\'as, S.~Janson, and O.~Riordan.
\newblock The phase transition in inhomogeneous random graphs.
\newblock {\em Random Structures \& Algorithms}, {\bfseries 31}(1), 3--122
  (2007).

\bibitem{Decelle_SBM_long}
A.~Decelle, F.~Krzakala, C.~Moore, and L.~Zdeborov\'a.
\newblock Asymptotic analysis of the stochastic block model for modular
  networks and its algorithmic applications.
\newblock {\em Phys. Rev. E}, {\bfseries 84}, 066106 (2011).

\bibitem{Cris_review_SBM}
C.~Moore.
\newblock The Computer Science and Physics of Community Detection: Landscapes,
  Phase Transitions, and Hardness.
\newblock {\em Bulletin of EATCS}, {\bfseries 1}(121) (2017).

\bibitem{Abbe_review_SBM}
E.~Abbe.
\newblock Community Detection and Stochastic Block Models: Recent Developments.
\newblock {\em Journal of Machine Learning Research}, {\bfseries 18}(177),
  1--86 (2018).

\bibitem{BuAlBoPo16}
J.~Bun, R.~Allez, J.-P. Bouchaud, and M.~Potters.
\newblock Rotational Invariant Estimator for General Noisy Matrices.
\newblock {\em IEEE Transactions on Information Theory}, {\bfseries 62}(12),
  7475--7490 (2016).

\bibitem{MaKrMeZd22}
A.~Maillard, F.~Krzakala, M.~M\'ezard, and L.~Zdeborov\'a.
\newblock Perturbative construction of mean-field equations in extensive-rank
  matrix factorization and denoising.
\newblock {\em Journal of Statistical Mechanics: Theory and Experiment},
  {\bfseries 2022}(8), 083301 (2022).

\bibitem{BaMa22}
J.~Barbier and N.~Macris.
\newblock Statistical limits of dictionary learning: Random matrix theory and
  the spectral replica method.
\newblock {\em Phys. Rev. E}, {\bfseries 106}, 024136 (2022).

\bibitem{TrErKrMaZd22}
E.~Troiani, V.~Erba, F.~Krzakala, A.~Maillard, and L.~Zdeborov\'a.
\newblock Optimal denoising of rotationally invariant rectangular matrices.
\newblock {\em Proceedings of Mathematical and Scientific Machine Learning
  (MSML), PMLR}, {\bfseries 190}, 97--112 (2022).

\bibitem{PoBaMa23}
F.~Pourkamali, J.~Barbier, and N.~Macris.
\newblock Matrix Inference in Growing Rank Regimes.
\newblock {\em arXiv:2306.01412},  (2023).

\bibitem{Se24}
G.~Semerjian.
\newblock Matrix denoising: Bayes-optimal estimators via low-degree
  polynomials.
\newblock {\em J. Stat. Phys.}, {\bfseries 191}, 139 (2024).

\bibitem{BaCaKoOk24}
J.~Barbier, F.~Camilli, J.~Ko, and K.~Okajima.
\newblock On the phase diagram of extensive-rank symmetric matrix denoising
  beyond rotational invariance.
\newblock {\em arXiv:2411.01974},  (2024).

\bibitem{OlFi96}
B.~Olshausen and D.~Field.
\newblock Emergence of simple-cell receptive field properties by learning a
  sparse code for natural images.
\newblock {\em Nature}, {\bfseries 381}, 607--609 (1996).

\bibitem{ZiPe01}
M.~Zibulevsky and B.~A. Pearlmutter.
\newblock Blind source separation by sparse decomposition in a signal
  dictionary.
\newblock {\em Neural computation}, {\bfseries 13}(4), 863--882 (2001).

\bibitem{KrMuRaEnLeSe03}
K.~Kreutz-Delgado, J.~F. Murray, B.~D. Rao, K.~Engan, T.-W. Lee, and T.~J.
  Sejnowski.
\newblock Dictionary learning algorithms for sparse representation.
\newblock {\em Neural computation}, {\bfseries 15}(2), 349--396 (2003).

\bibitem{SpWaWr12}
D.~A. Spielman, H.~Wang, and J.~Wright.
\newblock Exact Recovery of Sparsely-Used Dictionaries.
\newblock In {\em Proceedings of the 25th Annual Conference on Learning
  Theory}, volume~23, pages 37.1--37.18, 2012.

\bibitem{RaFl12}
S.~Rangan and A.~K. Fletcher.
\newblock Iterative estimation of constrained rank-one matrices in noise.
\newblock In {\em 2012 IEEE International Symposium on Information Theory
  Proceedings}, pages 1246--1250, 2012.

\bibitem{BaDiMaKrLeZd16}
J.~Barbier, M.~Dia, N.~Macris, F.~Krzakala, T.~Lesieur, and L.~Zdeborov\'a.
\newblock {Mutual information for symmetric rank-one matrix estimation: A proof
  of the replica formula}.
\newblock {\em Advances in Neural Information Processing Systems 29 (NIPS
  2016)}, pages 424--432 (2016).

\bibitem{LeKrZd17}
T.~Lesieur, F.~Krzakala, and L.~Zdeborov\'a.
\newblock Constrained low-rank matrix estimation: phase transitions,
  approximate message passing and applications.
\newblock {\em Journal of Statistical Mechanics: Theory and Experiment},
  {\bfseries 2017}(7), 073403 (2017).

\bibitem{LeMi19}
M.~Lelarge and L.~Miolane.
\newblock Fundamental limits of symmetric low-rank matrix estimation.
\newblock {\em Probab. Theory Relat. Fields}, {\bfseries 173}, 859--929 (2019).

\bibitem{BaMa19}
J.~Barbier and N.~Macris.
\newblock The adaptive interpolation method: a simple scheme to prove replica
  formulas in Bayesian inference.
\newblock {\em Probab. Theory Relat. Fields}, {\bfseries 174}, 1133–1185
  (2019).

\bibitem{MoWe22}
A.~Montanari and A.~S. Wein.
\newblock Equivalence of Approximate Message Passing and Low-Degree Polynomials
  in Rank-One Matrix Estimation.
\newblock {\em arXiv:2212.06996},  (2022).

\bibitem{SaKa13}
A.~Sakata and Y.~Kabashima.
\newblock Statistical mechanics of dictionary learning.
\newblock {\em Europhysics Letters}, {\bfseries 103}(2), 28008 (2013).

\bibitem{KaKrMeSaZd16}
Y.~Kabashima, F.~Krzakala, M.~M\'ezard, A.~Sakata, and L.~Zdeborov\'a.
\newblock Phase Transitions and Sample Complexity in Bayes-Optimal Matrix
  Factorization.
\newblock {\em IEEE Transactions on Information Theory}, {\bfseries 62}(7),
  4228--4265 (2016).

\bibitem{Sc18}
H.~C. Schmidt.
\newblock Statistical Physics of Sparse and Dense Models in Optimization and
  Inference.
\newblock In {\em PhD thesis}, 2018.

\bibitem{BaKoRa24}
J.~Barbier, J.~Ko, and A.~A. Rahman.
\newblock A multiscale cavity method for sublinear-rank symmetric matrix
  factorization.
\newblock {\em arXiv:2403.07189},  (2024).

\bibitem{PoMa23}
F.~Pourkamali and N.~Macris.
\newblock Bayesian extensive-rank matrix factorization with rotational
  invariant priors.
\newblock In {\em Proceedings of the 37th International Conference on Neural
  Information Processing Systems}, NIPS '23, pages 24025--24073, 2024.

\bibitem{PoMa23b}
F.~Pourkamali and N.~Macris.
\newblock Rectangular Rotational Invariant Estimator for General Additive Noise
  Matrices.
\newblock {\em arXiv:2304.12264},  (2023).

\bibitem{LaMeGa23}
I.~D. Landau, G.~C. Mel, and S.~Ganguli.
\newblock Singular vectors of sums of rectangular random matrices and optimal
  estimation of high-rank signals: The extensive spike model.
\newblock {\em Phys. Rev. E}, {\bfseries 108}, 054129 (2023).

\bibitem{CaMe23a}
F.~Camilli and M.~M\'ezard.
\newblock Matrix factorization with neural networks.
\newblock {\em Phys. Rev. E}, {\bfseries 107}, 064308 (2023).

\bibitem{CaMe23b}
F.~Camilli and M.~M\'ezard.
\newblock The decimation scheme for symmetric matrix factorization.
\newblock {\em Journal of Physics A: Mathematical and Theoretical}, {\bfseries
  57}(8), 085002 (2024).

\bibitem{Hastie_book}
T.~Hastie, R.~Tibshirani, and J.~Friedman.
\newblock {\em The elements of statistical learning, 2d edition}.
\newblock Springer, 2009.

\bibitem{BaDyLiPo24}
L.~P. Barnes, A.~Dytso, J.~Liu, and H.~V. Poor.
\newblock Multivariate Priors and the Linearity of Optimal Bayesian Estimators
  under Gaussian Noise.
\newblock {\em arXiv:2401.16701},  (2024).

\bibitem{McDo98}
I.~G. Macdonald.
\newblock {\em Symmetric functions and Hall polynomials}.
\newblock Oxford University Press, 1998.

\bibitem{We66}
H.~Weyl.
\newblock {\em The Classical Groups: Their Invariants and Representations}.
\newblock Princeton University Press, 1966.

\bibitem{ScWe22}
T.~Schramm and A.~S. Wein.
\newblock {Computational barriers to estimation from low-degree polynomials}.
\newblock {\em The Annals of Statistics}, {\bfseries 50}(3), 1833 -- 1858
  (2022).

\bibitem{HoSt17}
S.~B. Hopkins and D.~Steurer.
\newblock Efficient Bayesian Estimation from Few Samples: Community Detection
  and Related Problems.
\newblock In {\em 2017 IEEE 58th Annual Symposium on Foundations of Computer
  Science (FOCS)}, pages 379--390, 2017.

\bibitem{KuWeBa19}
D.~Kunisky, A.~S. Wein, and A.~S. Bandeira.
\newblock Notes on Computational Hardness of Hypothesis Testing: Predictions
  Using the Low-Degree Likelihood Ratio.
\newblock In {\em Mathematical Analysis, its Applications and Computation},
  pages 1--50. Springer International Publishing, 2022.

\bibitem{KuMoWe24}
D.~Kunisky, C.~Moore, and A.~S. Wein.
\newblock Tensor cumulants for statistical inference on invariant
  distributions.
\newblock {\em arXiv:2404.18735},  (2024).

\bibitem{MeMiMo21}
S.~Mei, T.~Misiakiewicz, and A.~Montanari.
\newblock Learning with invariances in random features and kernel models.
\newblock In M.~Belkin and S.~Kpotufe, editors, {\em Proceedings of Thirty
  Fourth Conference on Learning Theory}, volume 134 of {\em Proceedings of
  Machine Learning Research}, pages 3351--3418. PMLR, 15--19 Aug 2021.

\bibitem{AbPeSi18}
E.~Abbe, J.~M. Pereira, and A.~Singer.
\newblock Estimation in the Group Action Channel.
\newblock In {\em 2018 IEEE International Symposium on Information Theory
  (ISIT)}, pages 561--565, 2018.

\bibitem{BaBlKiNiPeWe23}
A.~S. Bandeira, B.~Blum-Smith, J.~Kileel, J.~Niles-Weed, A.~Perry, and A.~S.
  Wein.
\newblock Estimation under group actions: Recovering orbits from invariants.
\newblock {\em Applied and Computational Harmonic Analysis}, {\bfseries 66},
  236--319 (2023).

\bibitem{BuDoLeMaWaWi24}
P.~B\"{u}rgisser, M.~L. Do\u{g}an, V.~Makam, M.~Walter, and A.~Wigderson.
\newblock {Complexity of Robust Orbit Problems for Torus Actions and the
  abc-Conjecture}.
\newblock In R.~Santhanam, editor, {\em 39th Computational Complexity
  Conference (CCC 2024)}, volume 300 of {\em Leibniz International Proceedings
  in Informatics (LIPIcs)}, pages 14:1--14:48, Dagstuhl, Germany, 2024. Schloss
  Dagstuhl -- Leibniz-Zentrum f{\"u}r Informatik.

\bibitem{AdDuLeMc84}
M.~Adrian, J.~Dubochet, J.~Lepault, and A.~W. McDowall.
\newblock Cryo-electron microscopy of viruses.
\newblock {\em Nature}, {\bfseries 308}, 32 (1984).

\bibitem{SiSh11}
A.~Singer and Y.~Shkolnisky.
\newblock Three-Dimensional Structure Determination from Common Lines in
  Cryo-EM by Eigenvectors and Semidefinite Programming.
\newblock {\em SIAM Journal on Imaging Sciences}, {\bfseries 4}(2), 543--572
  (2011).

\bibitem{Sig98}
F.~Sigworth.
\newblock A Maximum-Likelihood Approach to Single-Particle Image Refinement.
\newblock {\em Journal of Structural Biology}, {\bfseries 122}(3), 328--339
  (1998).

\bibitem{HaLiPo34}
G.~Hardy, J.~Littlewood, and G.~P{\'o}lya.
\newblock {\em Inequalities}.
\newblock Cambridge Mathematical Library. Cambridge University Press, 1934.

\bibitem{EdKa75}
J.~Edmonds and R.~M. Karp.
\newblock Theoretical Improvements in Algorithmic Efficiency for Network Flow
  Problems.
\newblock {\em J. ACM}, {\bfseries 19}(2), 248–--264 (1972).

\bibitem{GuShVe05}
D.~Guo, S.~Shamai, and S.~Verdu.
\newblock Mutual information and minimum mean-square error in Gaussian
  channels.
\newblock {\em IEEE Transactions on Information Theory}, {\bfseries 51}(4),
  1261--1282 (2005).

\bibitem{AlBrRaZe11}
F.~Altarelli, A.~Braunstein, A.~Ramezanpour, and R.~Zecchina.
\newblock Stochastic optimization by message passing.
\newblock {\em Journal of Statistical Mechanics: Theory and Experiment},
  {\bfseries 2011}(11), P11009 (2011).

\bibitem{CaZd11}
M.~Castellana and L.~Zdeborov\'a.
\newblock Adversarial satisfiability problem.
\newblock {\em Journal of Statistical Mechanics: Theory and Experiment},
  {\bfseries 2011}(03), P03023 (2011).

\bibitem{BaShSh08}
M.~Bayati, D.~Shah, and M.~Sharma.
\newblock {Max-Product for Maximum Weight Matching: Convergence, Correctness,
  and LP Duality}.
\newblock {\em IEEE Trans. Inform. Theory}, {\bfseries 54}(3), 1241--1251
  (2008).

\bibitem{MeMoNg18}
S.~Mei, A.~Montanari, and P.-M. Nguyen.
\newblock A mean field view of the landscape of two-layer neural networks.
\newblock {\em Proceedings of the National Academy of Sciences}, {\bfseries
  115}(33), E7665--E7671 (2018).

\bibitem{SaBiCaKrUrZd20}
S.~Sarao~Mannelli, G.~Biroli, C.~Cammarota, F.~Krzakala, P.~Urbani, and
  L.~Zdeborov\'a.
\newblock Marvels and Pitfalls of the Langevin Algorithm in Noisy
  High-Dimensional Inference.
\newblock {\em Phys. Rev. X}, {\bfseries 10}, 011057 (2020).

\bibitem{BeGhJa21}
G.~Ben~Arous, R.~Gheissari, and A.~Jagannath.
\newblock Online stochastic gradient descent on non-convex losses from
  high-dimensional inference.
\newblock {\em Journal of Machine Learning Research}, {\bfseries 22}(106),
  1--51 (2021).

\bibitem{GlUr21}
G.~Gluch and R.~Urbanke.
\newblock Noether: The More Things Change, the More Stay the Same.
\newblock {\em arXiv:2104.05508},  (2021).

\bibitem{HaCh23}
K.~Hajjar and L.~Chizat.
\newblock On the symmetries in the dynamics of wide two-layer neural networks.
\newblock {\em Electronic Research Archive}, {\bfseries 31}(4), 2175--2212
  (2023).

\bibitem{DaTrArPeZdKr24}
Y.~Dandi, E.~Troiani, L.~Arnaboldi, L.~Pesce, L.~Zdeborov\'a, and F.~Krzakala.
\newblock The benefits of reusing batches for gradient descent in two-layer
  networks: breaking the curse of information and leap exponents.
\newblock In {\em Proceedings of the 41st International Conference on Machine
  Learning}, ICML'24. JMLR.org, 2025.

\end{thebibliography}

\end{document}